\documentclass[aps,prl,twocolumn,showpacs,groupedaddress,floatfix]{revtex4}

\usepackage{graphicx}
\usepackage{dcolumn}
\usepackage{bm}
\usepackage{amssymb}
\usepackage{multirow}
\usepackage{verbatim}
\usepackage{array}

\def\ifb{fb$^{-1}$}
\def\pp{$p\bar{p}$}
\def\tevE{$\sqrt{s}=1.96$~TeV}
\def\wjets{$W+$jets}
\def\zjets{$Z+$jets}
\def\ttbar{$t\bar{t}$}
\def\met{\ensuremath{E\kern-0.57em/_{T}}}

\begin{document}
  
  
  \title{Evidence of $WW+WZ$ production with lepton $+$ jets\\ final states in \pp\ collisions at \tevE} 
%
\author{V.M.~Abazov$^{36}$}
\author{B.~Abbott$^{75}$}
\author{M.~Abolins$^{65}$}
\author{B.S.~Acharya$^{29}$}
\author{M.~Adams$^{51}$}
\author{T.~Adams$^{49}$}
\author{E.~Aguilo$^{6}$}
\author{M.~Ahsan$^{59}$}
\author{G.D.~Alexeev$^{36}$}
\author{G.~Alkhazov$^{40}$}
% ...

\affiliation{\vspace{0.1 in}(The D\O\ Collaboration)\vspace{0.1 in}}
% ...
\affiliation{$^{6}$University of Alberta, Edmonton, Alberta, Canada,
                Simon Fraser University, Burnaby, British Columbia, Canada,
                York University, Toronto, Ontario, Canada, and
                McGill University, Montreal, Quebec, Canada}
% ...
\affiliation{$^{29}$Tata Institute of Fundamental Research, Mumbai, India}
% ...
\affiliation{$^{36}$Joint Institute for Nuclear Research, Dubna, Russia}
% ...
\affiliation{$^{40}$Petersburg Nuclear Physics Institute,
                St. Petersburg, Russia}
% ...
\affiliation{$^{49}$Florida State University, Tallahassee, Florida 32306, USA}
% ...
\affiliation{$^{51}$University of Illinois at Chicago,
                Chicago, Illinois 60607, USA}
% ...
\affiliation{$^{59}$Kansas State University, Manhattan, Kansas 66506, USA}
% ...
\affiliation{$^{65}$Michigan State University,
                East Lansing, Michigan 48824, USA}
% ...
\affiliation{$^{75}$University of Oklahoma, Norman, Oklahoma 73019, USA}
% ...

  \date{\today}
  
  \begin{abstract}

    We present first evidence for $WW+WZ$ production in lepton+jets
    final states at a hadron collider.  The data correspond to
    1.07~\ifb\ of integrated luminosity collected with the D0 detector
    at the Fermilab Tevatron in \pp\ collisions at \tevE.
    The observed cross section for $WW+WZ$ production is $20.2\pm
    4.5$~pb, consistent with the standard model and more precise than
    previous measurements in fully leptonic final states.  The probability 
    that background fluctuations alone produce this excess is
    $<5.4\times 10^{-6}$, which corresponds to a significance of 4.4
    standard deviations.

  \end{abstract}
  

  \maketitle

  The production of vector-boson pairs in $p\bar{p}$ collisions ($WW$,
  $WZ$, or $ZZ$) provides important tests of the electroweak sector of
  the standard model (SM). The next-to-leading-order (NLO) cross
  sections for $WW$ and $WZ$ production in $p\bar{p}$ collisions at
  $\sqrt{s}=1.96$ GeV predicted by the SM are $\sigma(WW)=12.4\pm
  0.8$~pb and $\sigma(WZ)= 3.7\pm 0.3$~pb~\cite{bib:Campbell}. A
  discrepancy with this expectation or deviations in the predicted
  kinematic distributions could signal the presence of new physics,
  e.g., originating from anomalous trilinear gauge boson
  couplings~\cite{bib:anocoups}.  The production of two weak bosons is
  also relevant to searches for the Higgs boson or for new particles
  in extensions of the SM. Production of $WW$ and $WZ$ in $p\bar{p}$
  collisions at the Fermilab Tevatron Collider has thus far been
  observed only in fully leptonic decay modes~\cite{bib:dz,bib:cdf}.
  Previous searches for $WW$ and $WZ$ in lepton+jets final
  states~\cite{bib:D0RunI,bib:CDFRunII}, which benefit from a higher
  branching ratio relative to fully leptonic channels, were hindered
  by large backgrounds from jets produced in association with a $W$
  boson (\wjets).

  In this Letter we report first evidence from a hadron collider for the
  production of a $W$ boson that decays leptonically, associated with
  a second vector boson $V$ ($V$=$W$ or $Z$) that decays into
  $q\bar{q}$ ($WV$$\rightarrow \ell\nu q\bar{q};\ \,\ell$$=$$e, \mu$).
  The limited dijet mass resolution ($\approx$ 18\% for dijets
  from $W/Z$ decays) results in a significant overlap of the 
  $W$$\rightarrow q\bar{q}$ and $Z$$\rightarrow q\bar{q}$ dijet mass peaks. We
  therefore consider $WW$ and $WZ$ simultaneously, assuming the ratio
  of their cross sections as predicted by the SM.  The use of improved
  multivariate event classification and new statistical
  techniques~\cite{bib:poisson}, as well as an increased integrated
  luminosity, make the $WV$ signal in lepton+jets final states more
  distinguishable from \wjets\ background and more accessible to
  measurement than in the past~\cite{bib:D0RunI,bib:CDFRunII}.  This
  analysis also provides a valuable proving ground for such advanced
  techniques, now ubiquitous in Higgs searches at the Tevatron.

  We analyze 1.07~\ifb\ of data collected with the D0
  detector~\cite{bib:detector} at a center-of-mass energy of 1.96~TeV
  at the Tevatron.  Candidate $e\nu q\bar{q}$ events must pass a
  trigger based on a single electron or electron+jet(s) requirement
  that has an efficiency of $98^{+2}_{-3}$\%.  A suite of triggers for
  $\mu\nu q\bar{q}$ candidate events achieves an efficiency of $>95\%$
  at 95\% confidence level.

  To select $WV$$\rightarrow \ell\nu q\bar{q}$ candidates, we require:
  a single reconstructed lepton (electron or muon)~\cite{bib:leptons}
  with transverse momentum $p_T>20$~GeV and pseudorapidity
  $|\eta|<1.1\ (2.0)$ for electrons (muons); the imbalance in
  transverse energy to be $\met>20$~GeV; and at least two
  jets~\cite{bib:JetCone} with $p_T>20$~GeV and $|\eta|<2.5$.  The jet
  of highest $p_T$ must have $p_T>30$~GeV. To reduce background from
  processes that do not contain $W$$\rightarrow \ell\nu$, we require a
  ``transverse'' mass~\cite{bib:smithUA1} of $M_T^{\ell
  \nu}>35$~GeV. The lepton must be spatially matched to a track
  reconstructed in the central tracker that originates from the
  primary vertex.  Electrons (muons) must be isolated from other
  particles in the calorimeter (and central tracker)~\cite{bib:iso}.


  Signal and background processes containing charged leptons are
  modeled via Monte Carlo (MC) simulation.  The signal includes all
  possible $W$ and $Z$ decays, including their decays to
  leptons.  The diboson signal (WW and WZ) is generated with {\sc
  pythia}~\cite{bib:PYTHIA} using \textsc{CTEQ6L} parton distribution
  functions (PDFs).  The fixed-order matrix element (FOME) generator
  {\sc alpgen}~\cite{bib:ALPGEN} with \textsc{CTEQ6L1} PDFs is used to
  generate \wjets, \zjets, and \ttbar\ events to leading order at the
  parton level.  The FOME generator {\sc comphep}~\cite{bib:CompHEP}
  is used to produce single top-quark MC samples.  {\sc alpgen} and
  {\sc comphep} are interfaced to {\sc pythia} for subsequent parton
  showering and hadronization.  All simulated events undergo a {\sc
  geant}-based~\cite{bib:GEANT} detector simulation and are
  reconstructed using the same programs as used for D0 data.  The MC
  samples are normalized using next-to-leading-order (NLO) or
  next-to-next-to-leading-order predictions for SM cross sections,
  except \wjets\, which is scaled to the data.

  The probability for multijet events with misidentified leptons to
  pass all selection requirements is small; however, because of the
  copious production of multijet events, the background from this
  source cannot be ignored.  For $\mu\nu q\bar{q}$, the multijet
  background is modeled with data that fail the muon isolation
  requirements, but pass all other selections. The normalization is
  determined from a fit to the $M_T^{\ell\nu}$ distribution.  For
  $e\nu q\bar{q}$, the multijet background is estimated using a
  ``loose-but-not-tight'' data sample obtained by selecting events
  that pass loosened electron quality requirements, but fail the tight
  electron quality criteria~\cite{bib:leptons}.  This sample is
  normalized by the probability for a jet that passes the ``loose''
  electron requirements to also pass the tight requirement. Both
  $\mu\nu q\bar{q}$ and $e\nu q\bar{q}$ multijet samples are corrected
  for contributions from all processes modeled through MC.

  Accurate modeling of the selected events is vital.  The dominant
  background is \wjets, and the modeling of {\sc alpgen} \wjets\ and
  sources of uncertainty are therefore studied in great detail.
  Comparison of {\sc alpgen} with other generators and with data shows
  discrepancies~\cite{bib:ALPGENcomp} in jet $\eta$ and dijet angular
  separation.  Data are used to correct these quantities in the {\sc
  alpgen} $W+$jets and $Z+$jets samples.  The possible bias in this
  procedure from the presence of the diboson signal in data is small,
  but is nevertheless taken into account via a systematic uncertainty.
  Systematic effects on the differential distributions of the {\sc
  alpgen} $W$+jets and $Z$+jets MC events from changes of the
  renormalization and factorization scales and of the parameters used
  in the MLM parton-jet matching algorithm~\cite{bib:MLM} are also
  considered.  Uncertainties on PDFs, as well as uncertainties from
  object reconstruction and identification, are evaluated for all MC
  samples.  We consider the effect of systematic uncertainty both on
  the normalization and on the shape of differential distributions for
  signal and backgrounds~\cite{bib:EPAPS}.

  The signal and the backgrounds are further separated using a
  multivariate classifier to combine information from several
  kinematic variables.  This analysis uses a Random Forest (RF)
  classifier~\cite{bib:SPR1,bib:SPR2}.  Thirteen well-modeled
  kinematic variables~\cite{bib:EPAPS} that demonstrate a difference
  in probability density between signal and at least one of the
  backgrounds, such as dijet mass and \met, are used as inputs to the
  RF.  The RF is trained using half of each MC sample.  The other
  halves, along with the multijet background samples, are then
  evaluated by the RF and used in the measurement.


  \begin{figure}[tbp] 
    \begin{centering}
      \begin{tabular}{cc}
      \multirow{1}{*}[1.0in]{(a)} & \includegraphics[height=2.0in]{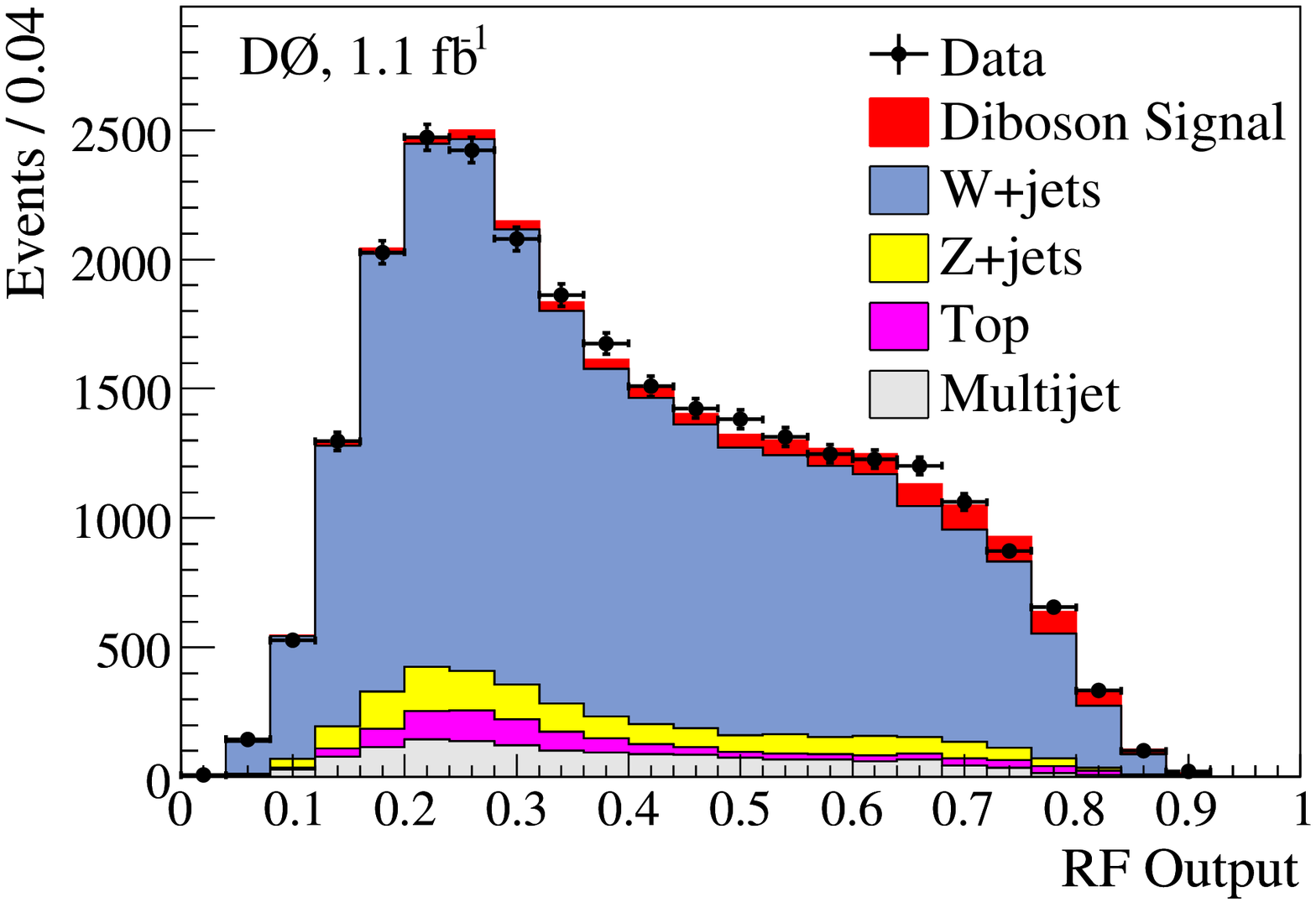}\\
      \multirow{1}{*}[1.0in]{(b)} & \includegraphics[height=2.0in]{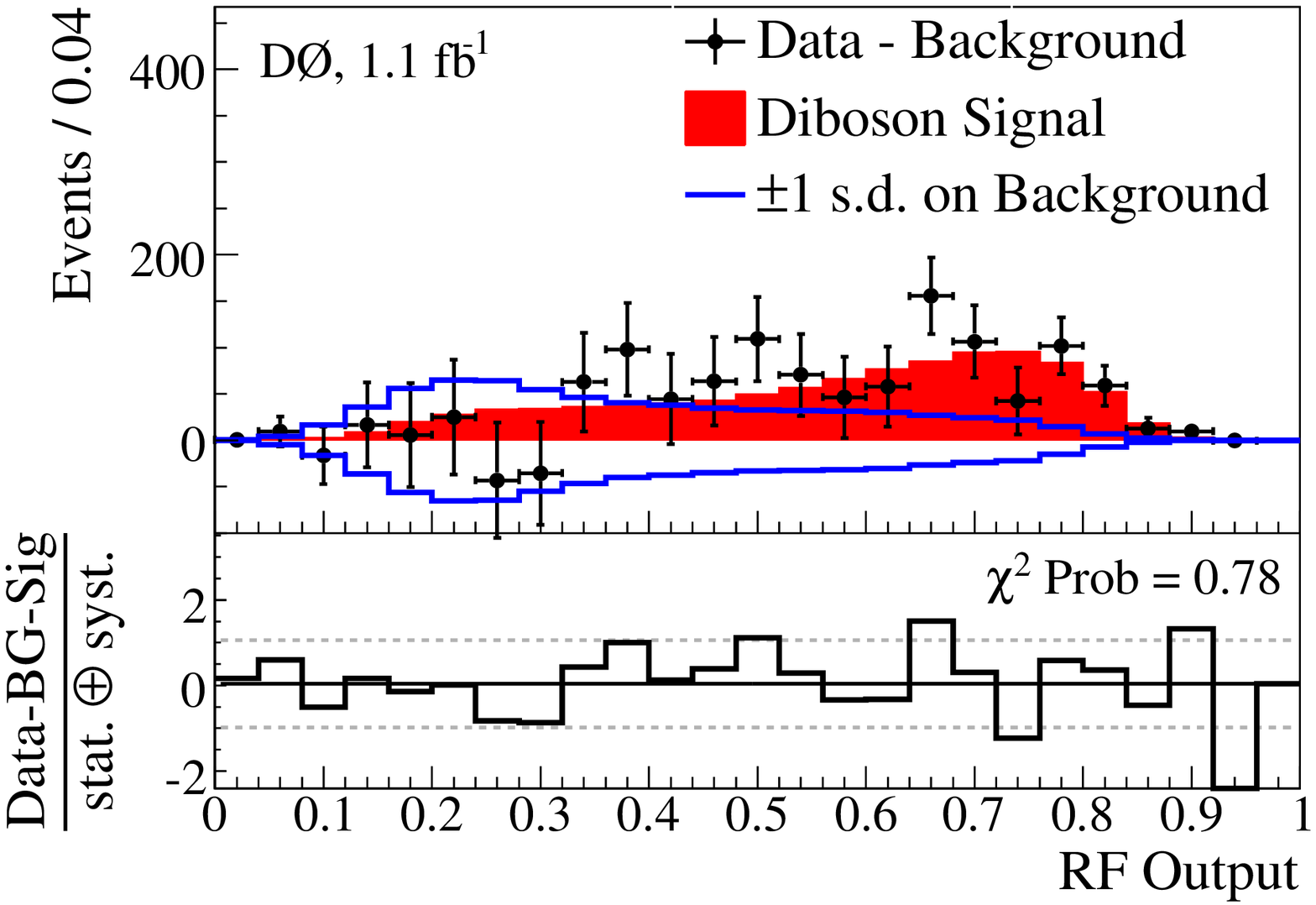}
    \end{tabular}
    
    \caption{(a) The RF output distribution from the combined $e\nu
    q\bar{q}$ and $\mu\nu q\bar{q}$ channels for data and MC
    predictions following the fit of MC to data. (b)~A comparison of
    the extracted signal (filled histogram) to background-subtracted
    data (points), along with the $\pm1$ standard deviation (s.d.) 
    systematic uncertainty on the background. The residual distance
    between the data points and the extracted signal, divided by
    the total uncertainty, is given at the bottom.}

    \label{fig:Fig1} \end{centering}
  \end{figure}

  \begin{figure}[tbp] 
    \begin{centering}
    \begin{tabular}{cc}
      \multirow{1}{*}[1.0in]{(a)} & \includegraphics[height=2.0in]{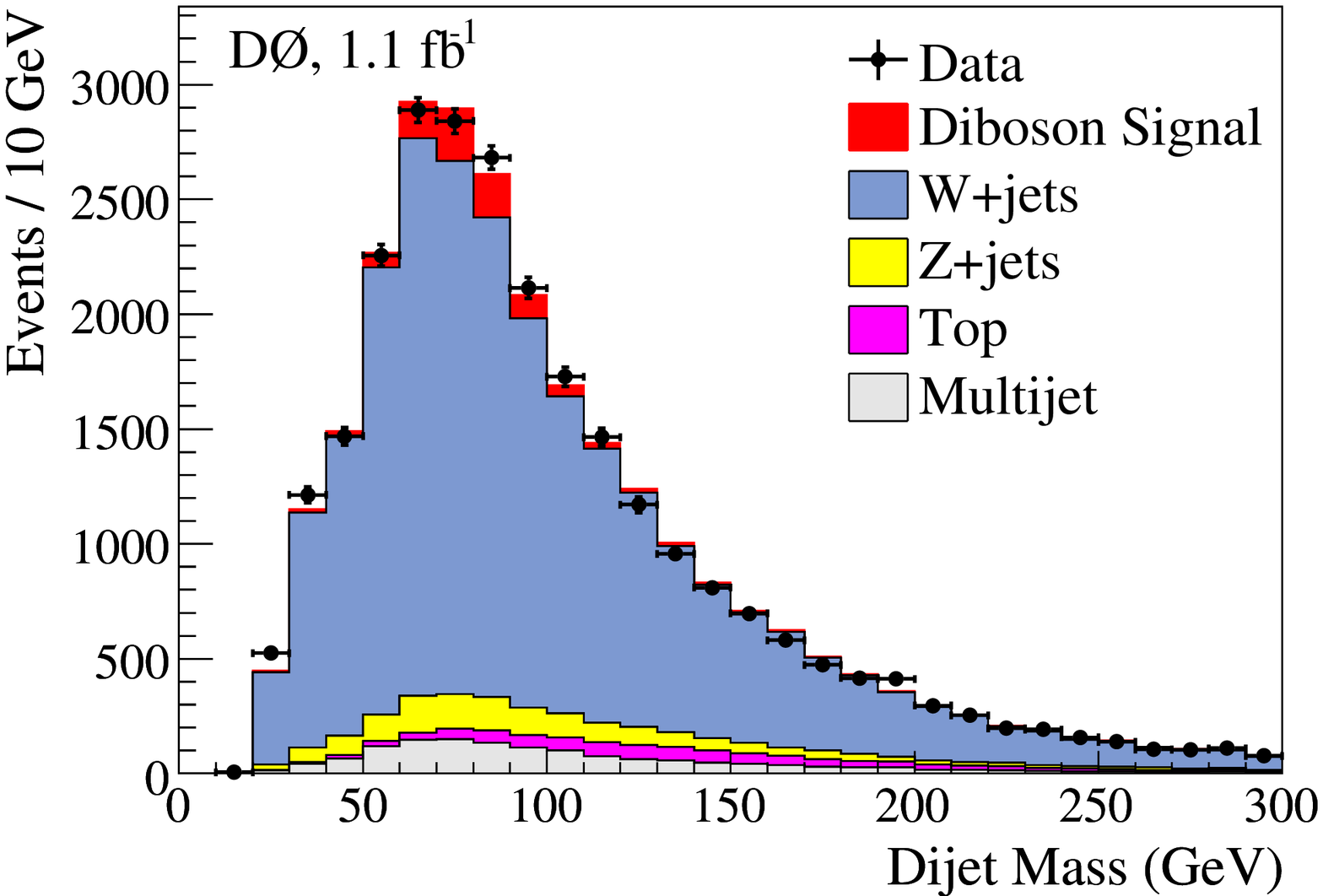}\\
      \multirow{1}{*}[1.0in]{(b)} & \includegraphics[height=2.0in]{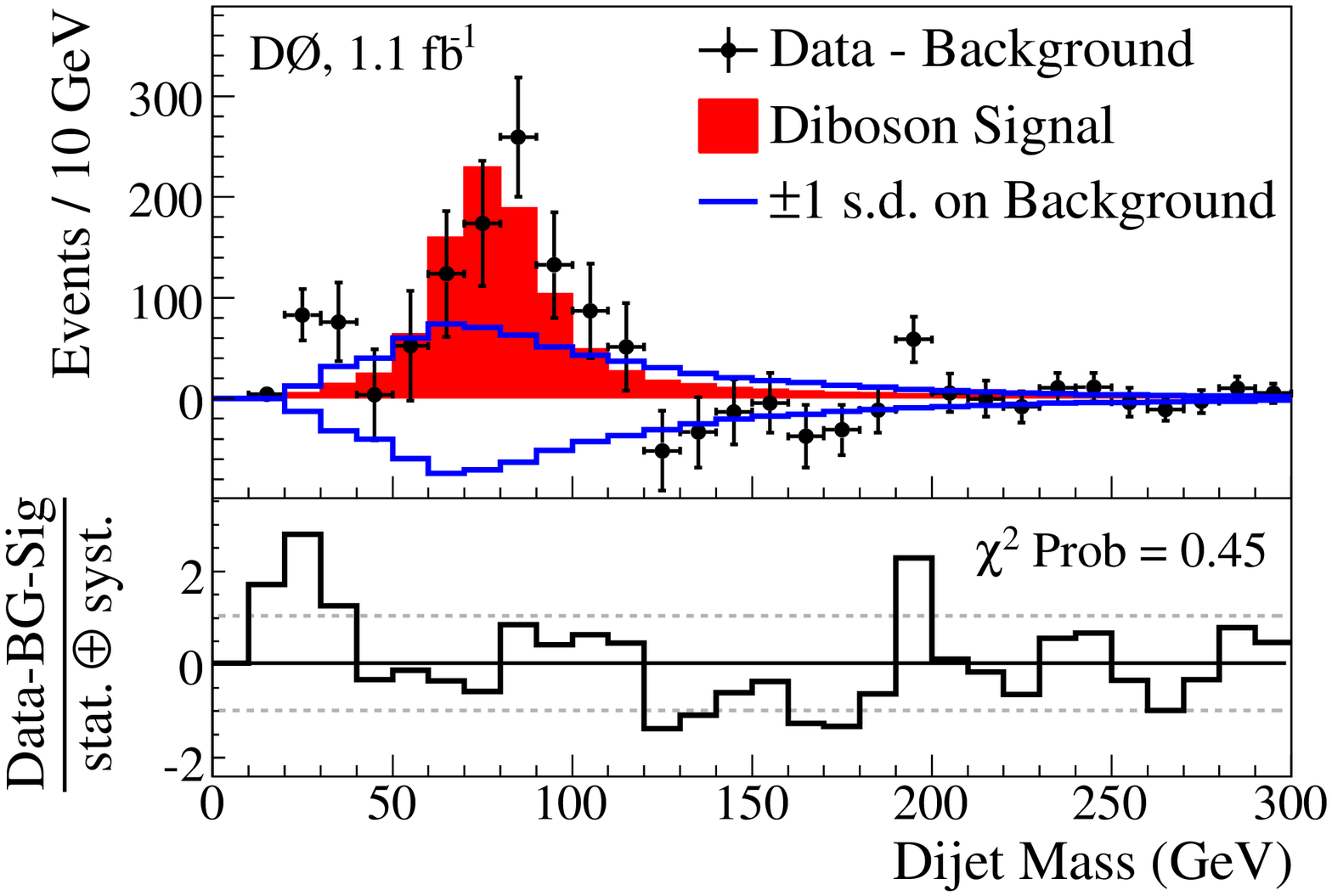}
    \end{tabular}
      
    \caption{(a) The dijet mass distribution from the combined $e\nu
    q\bar{q}$ and $\mu\nu q\bar{q}$ channels for data and MC
    predictions following the fit to the RF output. (b)~A comparison
    of the extracted signal (filled histogram) to
    background-subtracted data (points), along with the $\pm1$
    standard deviation (s.d.) systematic uncertainty on the
    background.The residual distance between the data points and the
    extracted signal, divided by the total uncertainty, is given
    at the bottom.}

      \label{fig:Fig2}
    \end{centering} 
  \end{figure}

  \begin{table}[tbp] 
  
    \caption{Measured number of events for signal and each background
    after the combined fit (with total uncertainties determined from
    the fit) and the number observed in data.}
    
    \label{tab:yields}
    \begin{ruledtabular}
      \begin{tabular}{l @{\extracolsep{\fill}} r @{$\ \pm\ $\extracolsep{0cm}} l @{\extracolsep{\fill}} r @{$\ \pm\ $\extracolsep{0cm}} l}
	           & \multicolumn{2}{c}{$e\nu q\bar{q}$ channel} & \multicolumn{2}{c}{$\mu\nu q\bar{q}$ channel} \\
	\hline	   
	Diboson signal        &   436  &  36  &   527  &  43 \\
	$W$+jets              & 10100  & 500  & 11910  & 590 \\
	$Z$+jets              &   387  &  61  &  1180  & 180 \\
	\ttbar\ + single top  &   436  &  57  &   426  &  54 \\
	Multijet              &  1100  & 200  &   328  &  83 \\
        \hline
	Total predicted       & 12460  & 550  &  14370 & 620 \\
	Data                  & \multicolumn{2}{c}{12473} & \multicolumn{2}{c}{14392} \\
      \end{tabular}
    \end{ruledtabular}
  \end{table}

  \begin{table*}[htb]

    \caption{ The signal cross section extracted from a simultaneous
    fit of the $WV$ cross section and the normalization factor for
    \wjets.  Also given are expected and observed p-values obtained by
    comparing the measurement with pseudo-experiments assuming no
    signal and the corresponding significance in number of standard
    deviations (s.d.) for a one-sided Gaussian integral.}

    \label{tab:xsecFit}
    \begin{ruledtabular}
      \begin{tabular}{lccc}
	Channel&Fitted signal $\sigma$ (pb)&Expected p-value (significance)&Observed p-value (significance)\\ 
	\hline
	 $e\nu q\bar{q}$ RF Output   & 18.0$\pm$3.7(stat)$\pm$5.2(sys)$\pm$1.1(lum) & $6.8\times10^{-3}$ (2.5 s.d.) & $3.2\times10^{-3}$ (2.7 s.d.) \\
         $\mu\nu q\bar{q}$ RF Output & 22.8$\pm$3.3(stat)$\pm$4.9(sys)$\pm$1.4(lum) & $1.8\times10^{-3}$ (2.9 s.d.) & $5.2\times10^{-5}$ (3.9 s.d.) \\
	 Combined RF Output          & 20.2$\pm$2.5(stat)$\pm$3.6(sys)$\pm$1.2(lum) & $1.5\times10^{-4}$ (3.6 s.d.) & $5.4\times10^{-6}$ (4.4 s.d.) \\
	 \hline
	 Combined Dijet Mass         & 18.5$\pm$2.8(stat)$\pm$4.9(sys)$\pm$1.1(lum) & $1.7\times10^{-3}$ (2.9 s.d.) & $4.4\times10^{-4}$ (3.3 s.d.) \\
      \end{tabular}
    \end{ruledtabular}
  \end{table*}

  The signal cross section is determined from a fit of signal and
  background RF templates to the data by minimizing a Poisson $\chi^2$
  function with respect to variations in the systematic
  uncertainties~\cite{bib:poisson}.  The magnitude of systematic
  uncertainties is effectively constrained by the regions of the RF
  distribution with low signal over background.  A Gaussian prior is
  used for each systematic uncertainty.  Different uncertainties are
  assumed to be mutually independent, but those common to multiple
  samples or lepton channels are assumed to be 100\% correlated.
  
  The fit simultaneously varies the $WV$ and \wjets\ contributions,
  thereby also determining the normalization factor for the \wjets\ MC
  sample.  This obviates the need for using the predicted {\sc alpgen}
  cross section, and provides a more rigorous approach that
  incorporates an unbiased uncertainty from \wjets\ when extracting
  the $WV$ cross section.  The normalization factor from the fit for
  the $W$+jets component is $1.53 \pm 0.13$, similar to the expected
  ratio of NLO to LO cross sections~\cite{bib:Campbell2}.  The
  measured yields for signal and each background are given in
  Table~\ref{tab:yields}.  Table~\ref{tab:xsecFit} contains the
  measured $WV$ cross section for each channel, separately and
  combined, showing consistent results between channels and the SM
  prediction of $\sigma(WV)=16.1\pm 0.9$~pb~\cite{bib:Campbell}.  The
  combined fit yields a cross section of
  20.2~$\pm$~2.5(stat)~$\pm$~3.6(sys)~$\pm$~1.2(lum)~pb.  The RF
  output distributions following the combined fit are shown in
  Fig.~\ref{fig:Fig1}, along with comparisons of consistency between
  the background-subtracted data and the extracted
  signal. Figure~\ref{fig:Fig2} shows analogous plots for the dijet
  mass after the combined fit to the RF output. The dominant
  systematic uncertainties arise from the modeling of the $W$+jets
  background and the jet energy scale, contributing 2.4 pb and 1.9 pb
  to the total systematic uncertainty~\cite{bib:EPAPS},
  respectively. The position of the dijet mass peaks in data and MC
  are consistent within one half standard deviation, which includes
  the relative data/MC uncertainty in energy scale. As a cross check,
  we also perform the measurement using only the dijet mass
  distribution.  The result, also given in Table~\ref{tab:xsecFit},
  although less precise, is consistent with that obtained using the RF
  output.

  The significance of the measurement is obtained via fits of the
  signal+background hypothesis to pseudo-data samples drawn from the
  background-only hypothesis~\cite{bib:singleTop}.  The observed (or
  expected) significance corresponds to the fraction of outcomes that
  yield a $WV$ cross section at least as large as that measured in
  data (as predicted by the SM).  The probabilities that background
  fluctuations could produce the expected and observed signal in each
  channel (p-values), separately and combined, are shown in
  Table~\ref{tab:xsecFit}, along with their corresponding significance
  (equivalent one-sided Gaussian probabilities).  The $\chi^2$ fit with
  respect to variations in the systematic
  uncertainties~\cite{bib:poisson} results in an improvement of the
  expected significance of the result from 2.4 (1.6) to 3.6 (2.9)
  standard deviations when using the RF output (dijet mass)
  discriminant.

  In summary, we measure $\sigma(WV)=20.2\pm 4.5$~pb (with $V$=$W$ or
  $Z$) in \pp\ collisions at \tevE. The probability that the
  backgrounds fluctuate to give an excess as large as observed in
  data is $<5.4\times10^{-6}$, corresponding to a significance of 4.4
  standard deviations.  This represents the first evidence for $WV$
  production in lepton+jets events at a hadron collider.  The result
  is more precise than previous independent measurements of $WW$ and
  $WZ$ yields in fully leptonic final states~\cite{bib:dz,bib:cdf} and
  consistent with the SM prediction of $\sigma(WV)=16.1\pm
  0.9$~pb~\cite{bib:Campbell}.  This work clearly demonstrates the
  ability of the D0 experiment to isolate a small signal in a large
  background in a final state of direct relevance to searches for a
  low mass Higgs, and thereby validates the analytical methods used in
  searches for Higgs bosons at the Tevatron~\cite{bib:tevcombo}.

%
We thank the staffs at Fermilab and collaborating institutions, 
and acknowledge support from the 
DOE and NSF (USA);
CEA and CNRS/IN2P3 (France);
FASI, Rosatom and RFBR (Russia);
CNPq, FAPERJ, FAPESP and FUNDUNESP (Brazil);
DAE and DST (India);
Colciencias (Colombia);
CONACyT (Mexico);
KRF and KOSEF (Korea);
CONICET and UBACyT (Argentina);
FOM (The Netherlands);
STFC (United Kingdom);
MSMT and GACR (Czech Republic);
CRC Program, CFI, NSERC and WestGrid Project (Canada);
BMBF and DFG (Germany);
SFI (Ireland);
The Swedish Research Council (Sweden);
CAS and CNSF (China);
and the
Alexander von Humboldt Foundation (Germany).

  \newpage
\def\mida{$+2/-3$}
\def\midb{$+0/-5$}
\def\sigmameas{$\sigma^{\mathrm{meas}}(WV)$}
\def\sigmath{$\sigma^{\mathrm{th}}(WV)$}



{\bf Supplemental Material: }

\section{Systematic Uncertainties}

Table~\ref{tab:systEMMU} gives the \% systematic uncertainties for
Monte Carlo simulations and multijet estimates.  We consider the
effect of systematic uncertainty both on the normalization and on the
shape of differential distributions for signal and
backgrounds. Although Table~\ref{tab:systEMMU} lists an uncertainty
for the $W$+jets simulation, this uncertainty is not used when
measuring the diboson signal cross section, for which the $W$+jets
normalization is a free parameter.  However, the size of the
uncertainty must be specified for generating the pseudo-data used in
the estimation of significance.  Also in the table is the contribution
of each systematic uncertainty to the total systematic uncertainty of
3.6~pb on the measured cross section, \sigmameas.  This total
systematic uncertainty is obtained from the systematic uncertainties
on the parameter in the fit to the Random Forest (RF) output,
\sigmameas/\sigmath, by multiplying each contribution by the theoretical cross
section \sigmath. The additional uncertainty on the integrated
luminosity for data (6.1\%) is therefore considered separately.
  
  \begin{table*}[htb] 

    \caption{The \% systematic uncertainties for Monte Carlo
    simulations and multijet estimates. Uncertainties are identical
    for both lepton channels except where otherwise indicated. The
    nature of the uncertainty, i.e., whether it refers to a
    differential dependence (D) or just normalization (N), is also
    provided. The values for uncertainties with a differential
    dependence correspond to the maximum amplitude of fluctuations in
    the RF output. Also provided is the contribution of each source to
    the total systematic uncertainty of 3.6~pb on the measured cross
    section, which does not include the additional uncertainty of
    6.1\% for the luminosity.  }
      
    \label{tab:systEMMU}
    \begin{ruledtabular}
      \begin{tabular}{l @{\extracolsep{\fill}} 
      r @{\ \ \extracolsep{\fill}} l @{\extracolsep{\fill}\ \ } 
      r @{\ \ \extracolsep{\fill}} l @{\extracolsep{\fill}\ \ } 
      r @{\ \ \extracolsep{\fill}} l @{\extracolsep{\fill}\ \ } 
      r @{\ \ \extracolsep{\fill}} l @{\extracolsep{\fill}\ \ } 
      r @{\ \ \extracolsep{\fill}} l @{\extracolsep{\fill}\ \ } 
      c
      c}
	\multicolumn{1}{c}{Source of systematic}
	& \multicolumn{2}{c}{\multirow{2}{*}{Diboson signal}}
	& \multicolumn{2}{c}{\multirow{2}{*}{$W$+jets}}
	& \multicolumn{2}{c}{\multirow{2}{*}{$Z$+jets}}
	& \multicolumn{2}{c}{\multirow{2}{*}{Top}}
	& \multicolumn{2}{c}{\multirow{2}{*}{Multijet}}
	& \multicolumn{1}{c}{\multirow{2}{*}{Nature}}
	& \multicolumn{1}{c}{\multirow{2}{*}{$\Delta\sigma\;(pb)$}}\\
	\multicolumn{1}{c}{uncertainty}
	&&
	&&
	&&
	&&
	&&
	&
	&\\

	\hline
        Trigger efficiency, $e\nu q\bar{q}$ channel          &&   \mida  &&  \mida   &&  \mida   &&  \mida   &&         & N& $<0.1$\\
        Trigger efficiency, $\mu\nu q\bar{q}$ channel        &&   \midb  &&  \midb   &&  \midb   &&  \midb   &&         & D& $<0.1$\\
	Lepton identification                                &&  $\pm$4  &&  $\pm$4  &&  $\pm$4  &&  $\pm$4  &&         & N& $<0.1$\\
	Jet identification                                   &&  $\pm$1  &&  $\pm$1  &&  $\pm$1  &&  $\pm<$1  &&        & D& 0.3\\
	Jet energy scale                                     &&  $\pm$4  &&  $\pm$9  &&  $\pm$9  &&  $\pm$4  &&         & D& 1.9\\
	Jet energy resolution                                &&  $\pm$3  &&  $\pm$4  &&  $\pm$4  &&  $\pm$4  &&         & N& $<0.1$\\

	Cross section                                        &&          && 

	                                                                    $\pm$20\footnote{The uncertainty on the cross section for $W$+jets is not used in
	the diboson signal cross section measurement (the $W$+jets
	normalization is a free parameter); however, it is needed
	for generating pseudo-data to estimate the significance of the observed signal.}

	                                                                             &&  $\pm$6  &&  $\pm$10 &&         & N& 1.1\\
	Multijet normalization, $e\nu q\bar{q}$ channel      &&          &&          &&          &&          && $\pm$20 & N& 0.9\\
	Multijet normalization, $\mu\nu q\bar{q}$ channel    &&          &&          &&          &&          && $\pm$30 & N& 0.5\\
	Multijet shape, $e\nu q\bar{q}$ channel              &&          &&          &&          &&          && $\pm$6  & D& $<0.1$\\
	Multijet shape, $\mu\nu q\bar{q}$ channel            &&          &&          &&          &&          && $\pm$10 & D& $<0.1$\\
	Diboson signal NLO/LO shape                          &&  $\pm$10 &&          &&          &&          &&         & D& $<0.1$\\
        Parton distribution function                         &&  $\pm$1  &&  $\pm$1  &&  $\pm$1  &&  $\pm$1  &&         & D& 0.2\\ 
	{\sc alpgen} $\eta$ and $\Delta R$ corrections       &&          &&  $\pm$1  &&  $\pm$1  &&          &&         & D& $<0.1$\\
	Renormalization and factorization scale              &&          &&  $\pm$3  &&  $\pm$3  &&          &&         & D& 0.9\\ 
	{\sc alpgen} parton-jet matching parameters          &&          &&  $\pm$4  &&  $\pm$4  &&          &&         & D& 2.4\\ 
      \end{tabular}
    \end{ruledtabular}
  \end{table*}

\section{Input Variables to the Random Forest Classifier}

The 13 kinematic variables used in the RF classifier are listed below,
and their distributions are shown in Fig.~\ref{fig:rf_inputs}.  The
variables are derived from characteristics of objects reconstructed
from observables in each event and can be loosely classified into
three categories: (i) variables based on the kinematics of individual
objects, (ii) variables based on the kinematics of multiple objects,
and (iii) variables based on the angular relationships among objects.
Several variables are calculated using the four-momentum of the dijet
system or the leptonic $W$ candidate ($W^{\ell \nu}$).  The
dijet system is defined as the four-momentum sum of the jets with
highest $p_T$ (jet$_1$) and second highest $p_T$ (jet$_2$).
$W^{\ell \nu}$ is reconstructed from the charged lepton and the
\met.  The neutrino from the $W$$\to \ell \nu$ decay is assigned the
transverse momentum defined by \met\  and a longitudinal momentum that
is calculated assuming the mass of the W for $\ell \nu$ ($M_W = 80.4$
GeV).  Of the two possible solutions, we choose the one that provides
the smaller total invariant mass of all objects in the event.

\begin{itemize}
  \item \textbf{Kinematics of Individual Objects:}
  \begin{enumerate}
   
  \item The imbalance in transverse energy (\met), which is defined by
  the imbalance in transverse momentum as determined from the summing
  of products of energies and cosines of polar angles of calorimeter
  cells relative to the center of the detector (corrected for
  transverse momenta of muons and energy scales for jets and electrons
  in the event).

  \item The jet with second highest $p_T$: $p_T(\mathrm{Jet}_2)$.

  \end{enumerate}

  \item \textbf{Kinematics of Multiple Objects:}
  \begin{enumerate}
  
  \item The ``transverse $W$ mass'' reconstructed from the charged
  lepton and the \met: $M_T^{\ell \nu} =
  \sqrt{2\;p_T^{\ell}\;\met\;(1-\cos(\Delta\phi(\ell,\met)))}$.

  \item The $p_T$ of the $W^{\ell \nu}$ candidate.

  \item The invariant mass of the dijet system.

  \item The magnitude of the leading jet momentum perpendicular to the
  plane of the dijet system:
  $\frac{|\vec{p}_T(\mathrm{Jet}_1+\mathrm{Jet}_2)\times\vec{p}_T(\mathrm{Jet}_1)|}{|\vec{p}_T(\mathrm{Jet}_1+\mathrm{Jet}_2)|}$,
  where ``$\times$'' represents the usual vector cross product.  This
  variable is calculated in the rest frame of the $W^{\ell \nu}$
  candidate and is denoted $p_{T}^{Rel}(\mathrm{Dijet}, \mathrm{Jet}_{1}))^{\mathrm{W Frame}}$.

  \item The magnitude of the second-leading jet momentum perpendicular
  to the plane of the dijet system:
  $\frac{|\vec{p}_T(\mathrm{Jet}_1+\mathrm{Jet}_2)\times\vec{p}_T(\mathrm{Jet}_2)|}{|\vec{p}_T(\mathrm{Jet}_1+\mathrm{Jet}_2)|}$.
  This variable is calculated in the laboratory frame and is denoted
  $p_{T}^{Rel}(\mathrm{Dijet}, \mathrm{Jet}_{2}))^{\mathrm{Lab Frame}}$.

  \item The angular separation between the two jets of highest $p_T$,
  weighted by the ratio of the transverse momentum of the
  second-leading jet and the $W^{\ell \nu}$ candidate: $\Delta
  R(\mathrm{Jet}_1,\mathrm{Jet}_2)\frac{
  p_T(\mathrm{Jet}_2)}{p_T(\ell)+\met}$. This variable is calculated
  in the rest frame of the $W^{\ell \nu}$ candidate and is denoted $k_{T}^{\mathrm{Min, W Frame}}$.

  \item The ``centrality'' of the charged lepton and jets system,
  defined as the scalar sum of transverse momenta divided by the sum
  of energies of the charged lepton and all jets in the event.

  \end{enumerate}

  \item \textbf{Angular Relationships of Objects:}
  \begin{enumerate}

  \item The azimuthal separation between the charged lepton and the
  \met\  vector: $\Delta \phi(\met, lepton)$.

  \item The cosine of the angle between the dijet system and the leading
  jet in the laboratory frame: $\cos(\angle(\mathrm{Dijet},\mathrm{Jet}_1))$.
    
  \item The cosine of the angle between the dijet system and the
  second-leading jet in the laboratory frame: $\cos(\angle(\mathrm{Dijet}, \mathrm{Jet}_2))$.
 
  \item Cosine of the angle between leading jet and the $W^{\ell \nu}$
  candidate: $\cos(\angle(W^{\ell \nu},\mathrm{Jet}_1))^{\mathrm{Dijet Frame}}$, evaluated in the rest frame of the dijet system.

  \end{enumerate}
\end{itemize}

\begin{figure*}[htb]
  \centering
  \includegraphics[width=5.9cm]{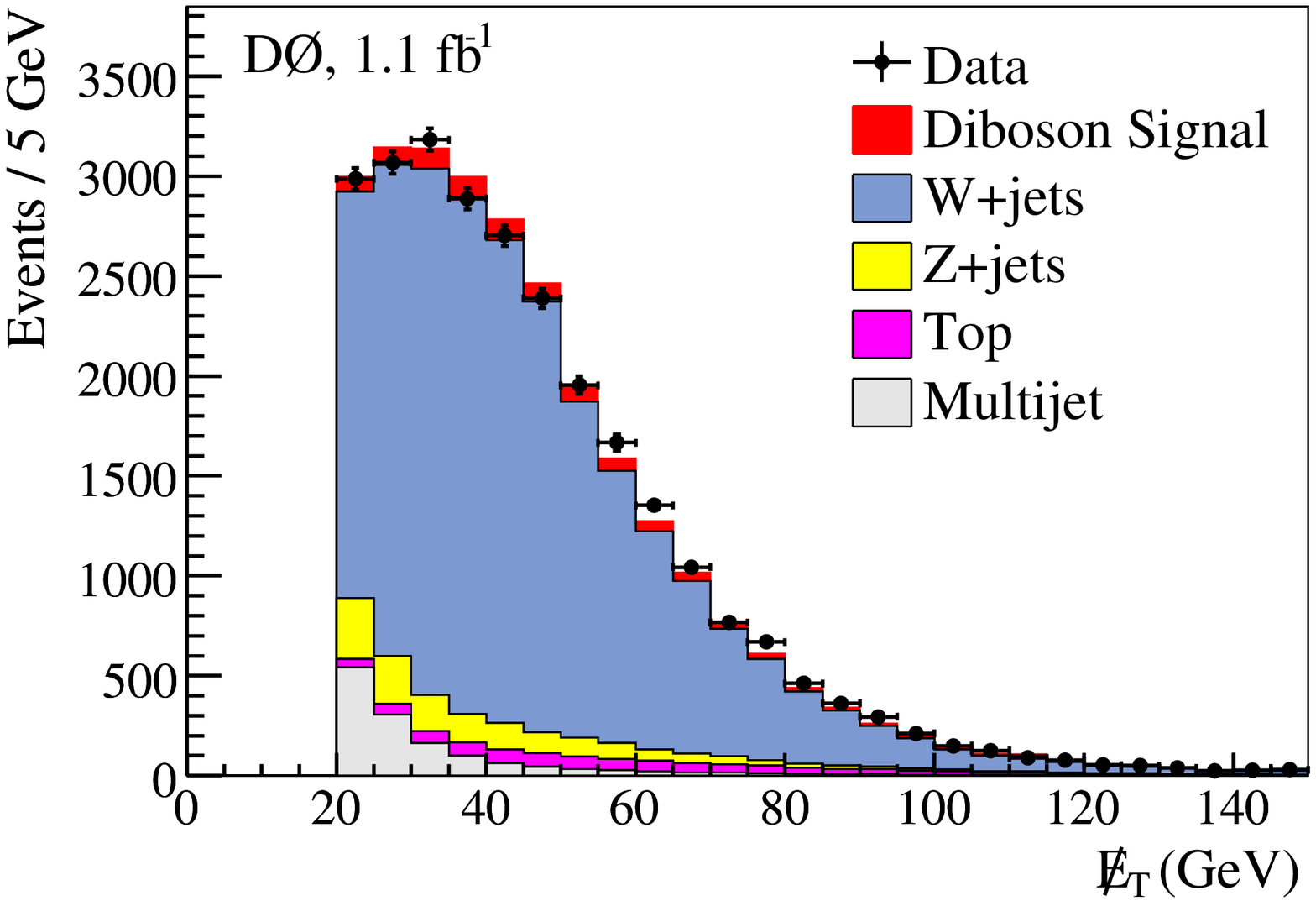}
  \includegraphics[width=5.9cm]{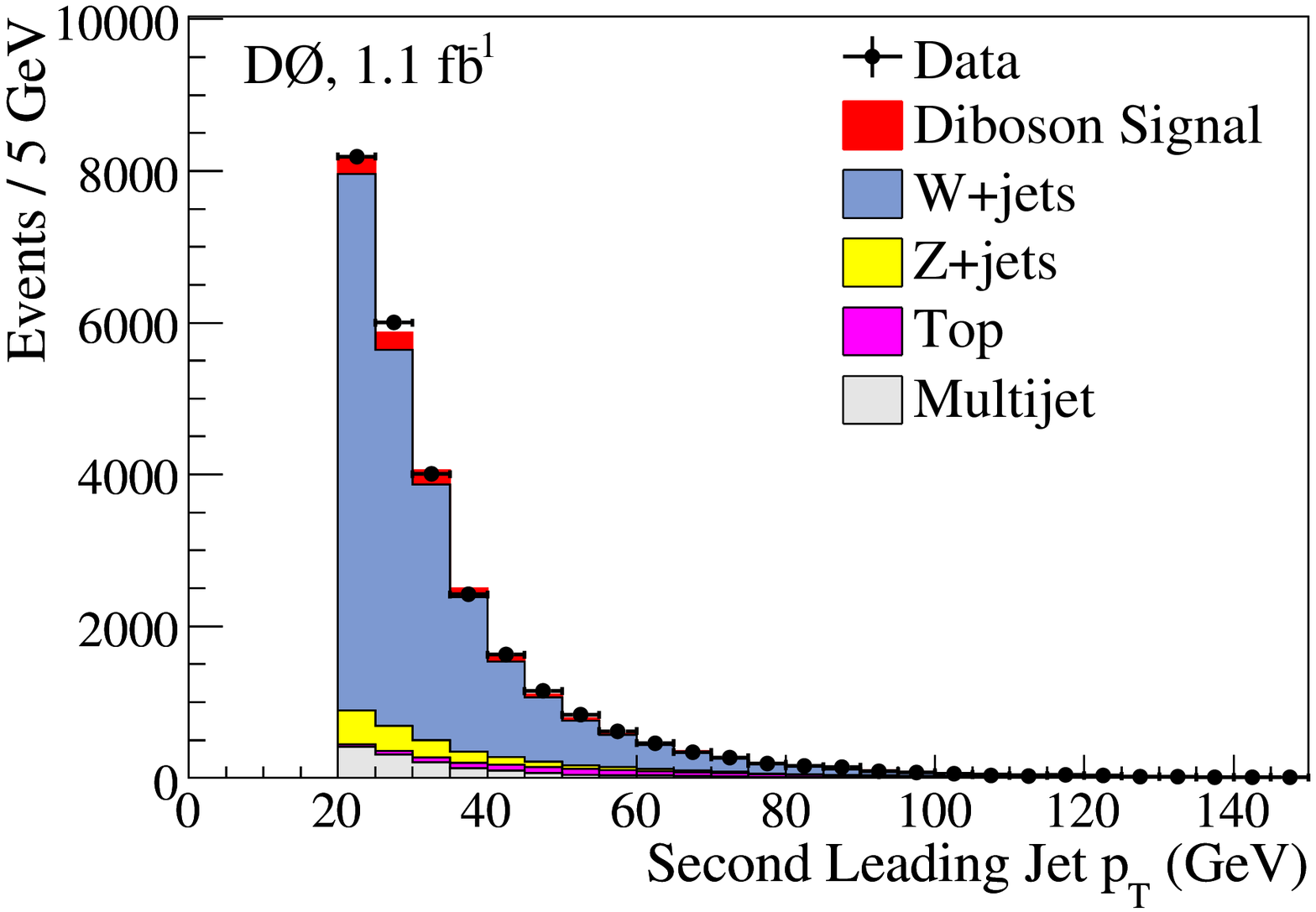}
  \includegraphics[width=5.9cm]{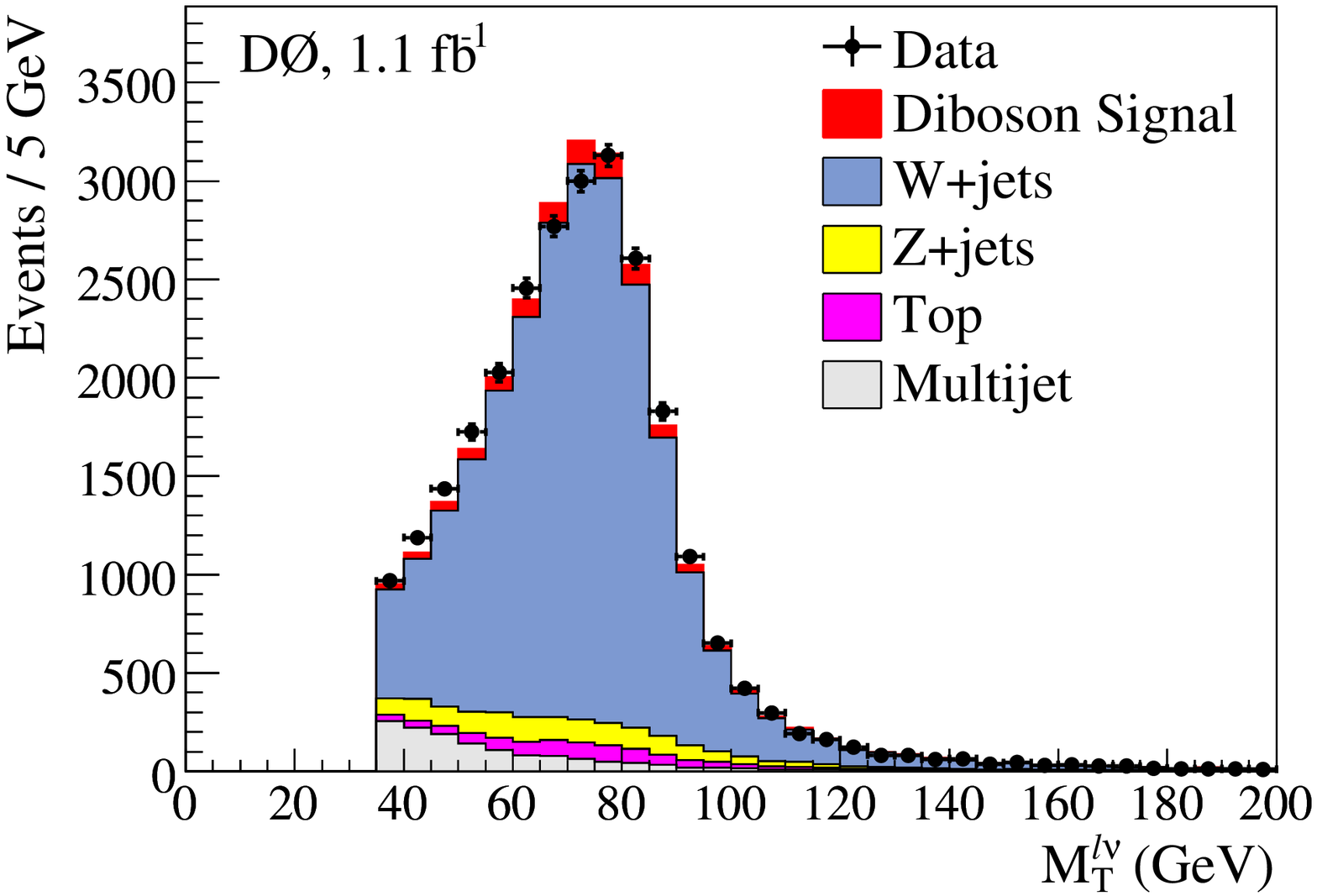}
  \includegraphics[width=5.9cm]{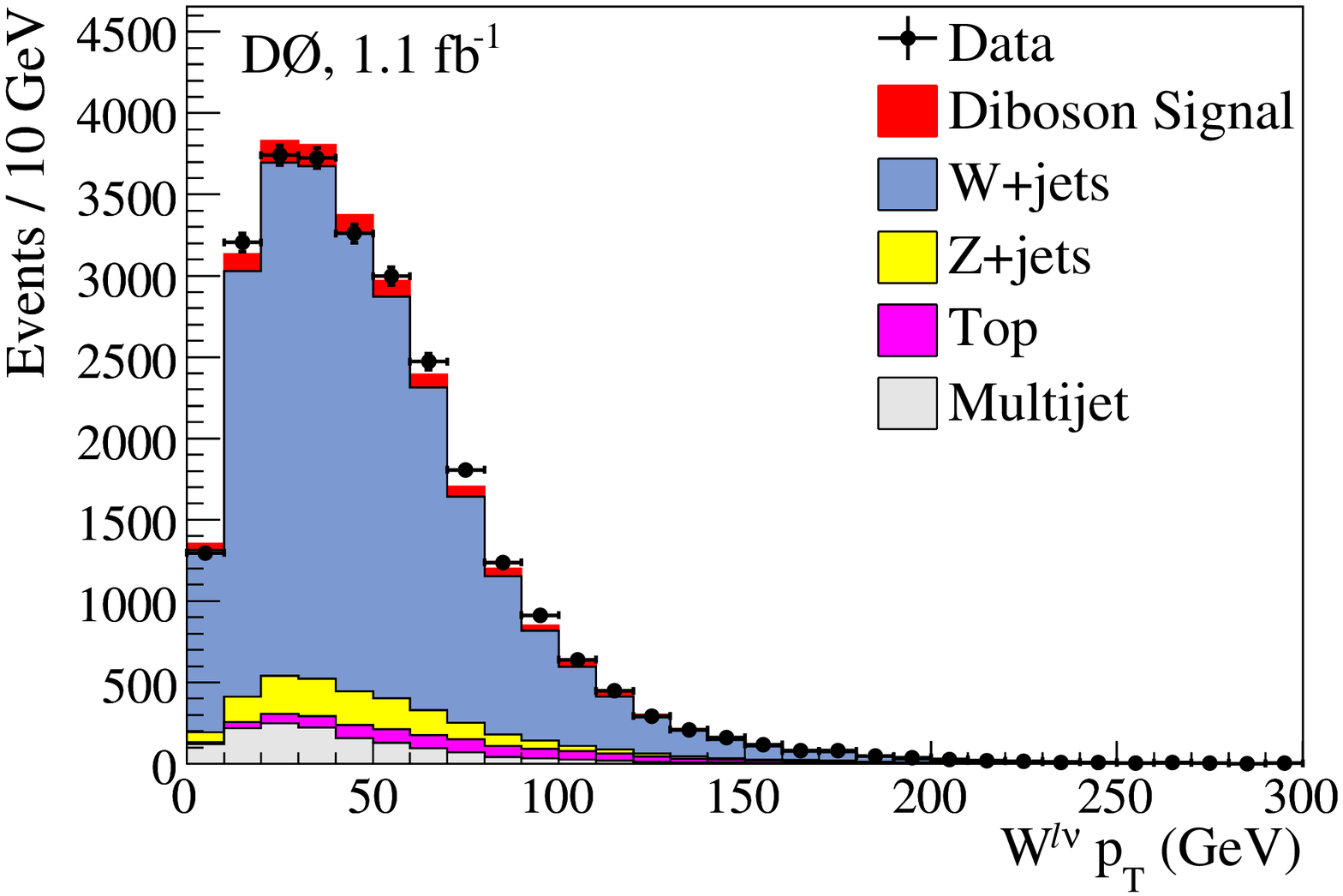}
  \includegraphics[width=5.9cm]{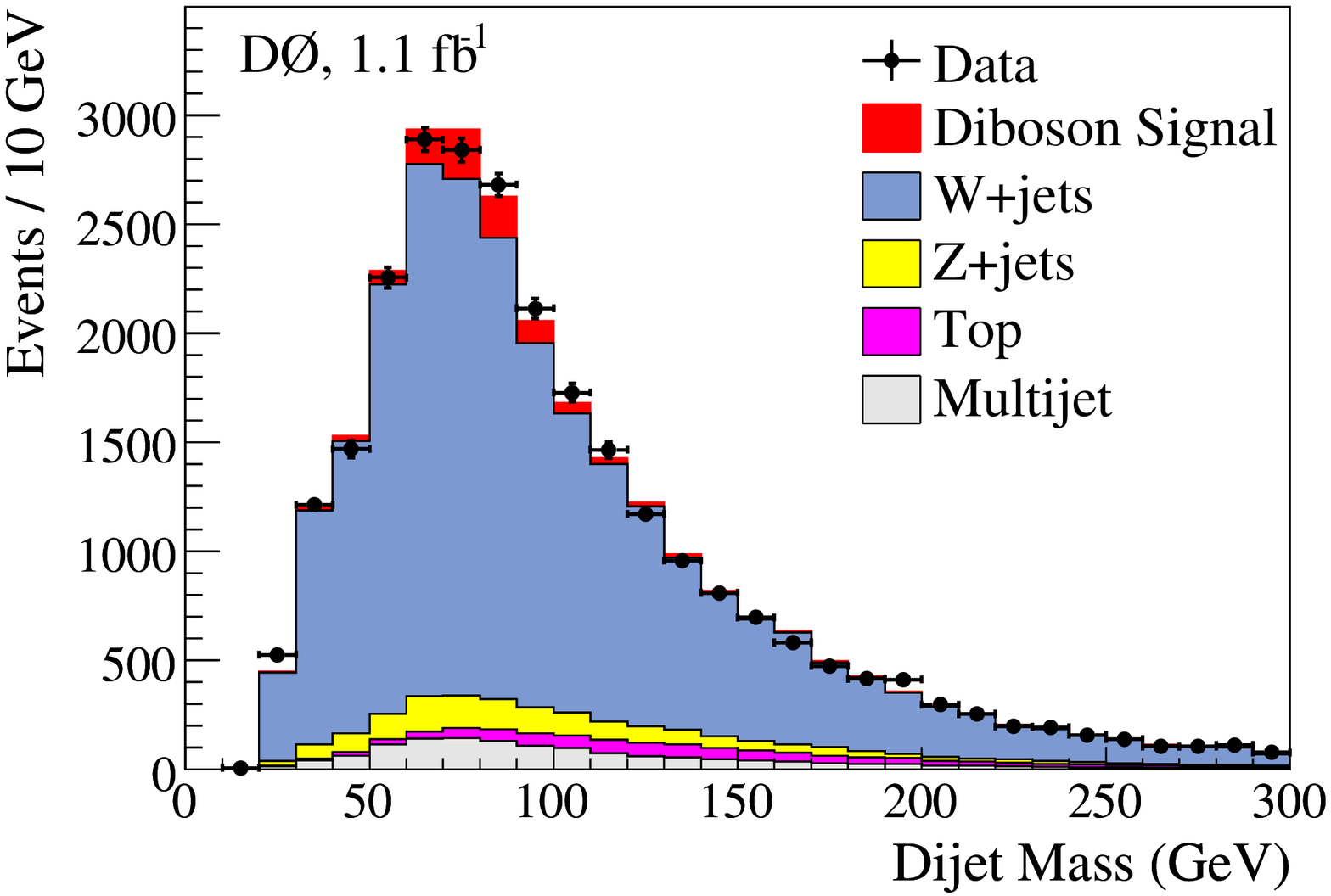}
  \includegraphics[width=5.9cm]{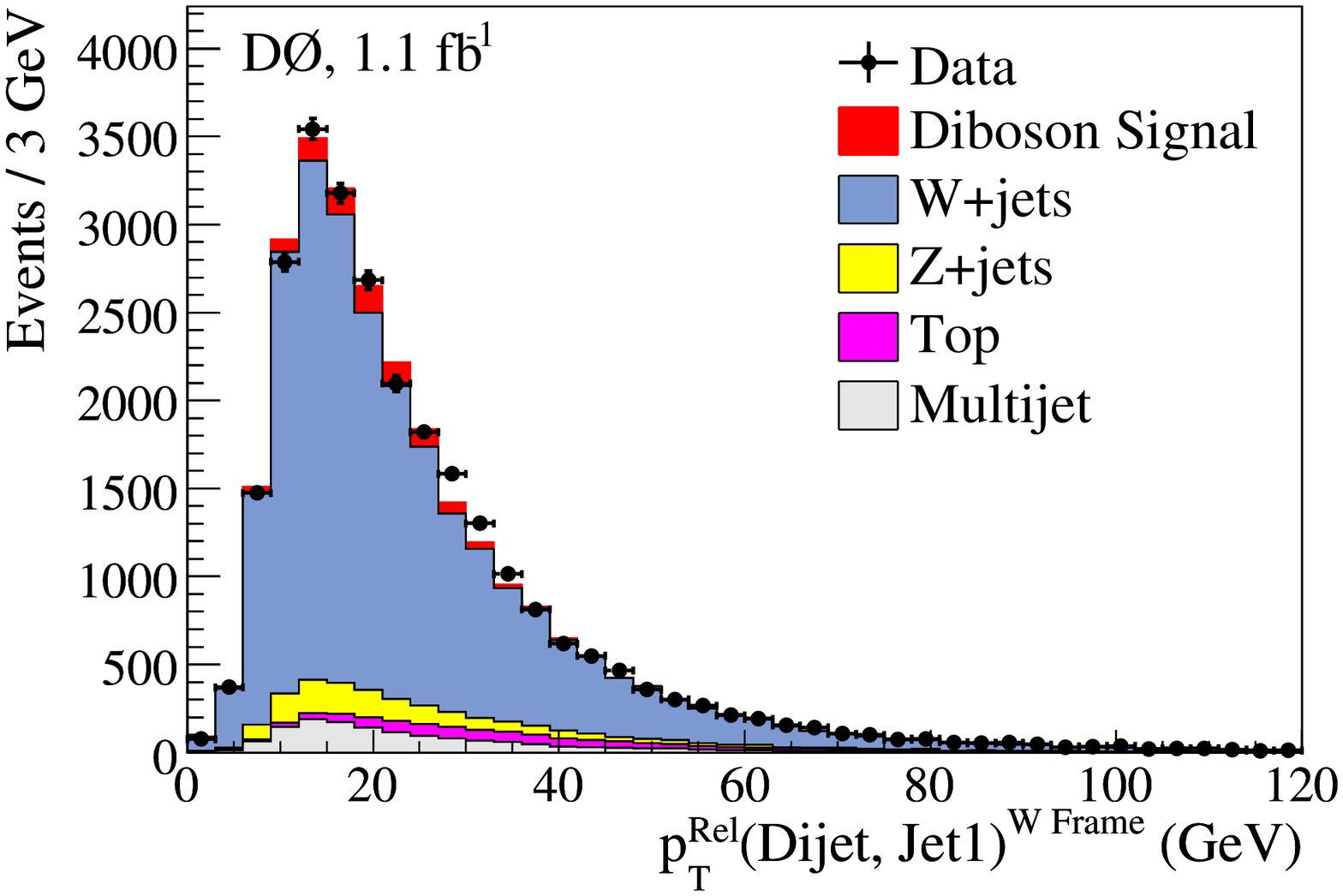}
  \includegraphics[width=5.9cm]{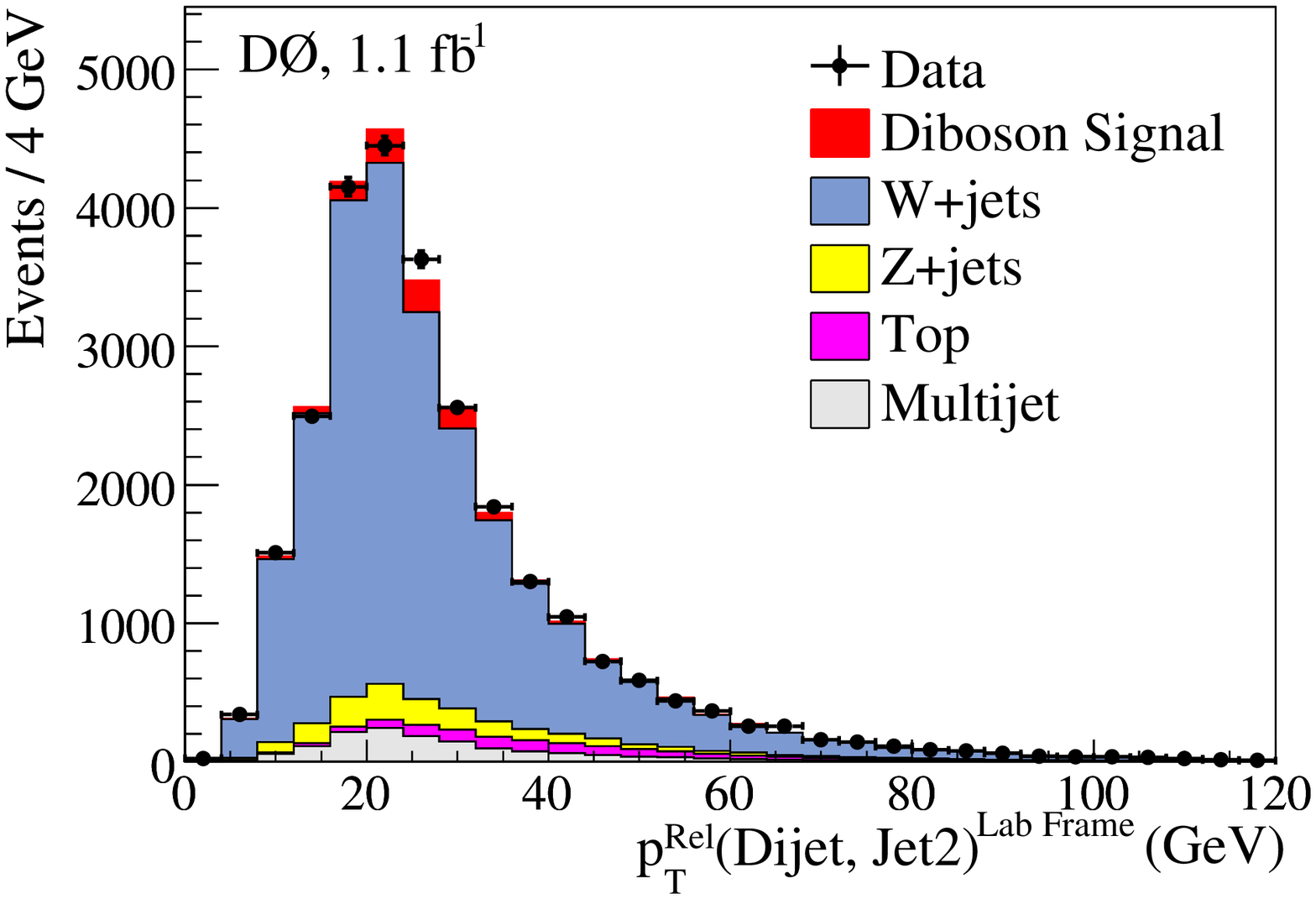}
  \includegraphics[width=5.9cm]{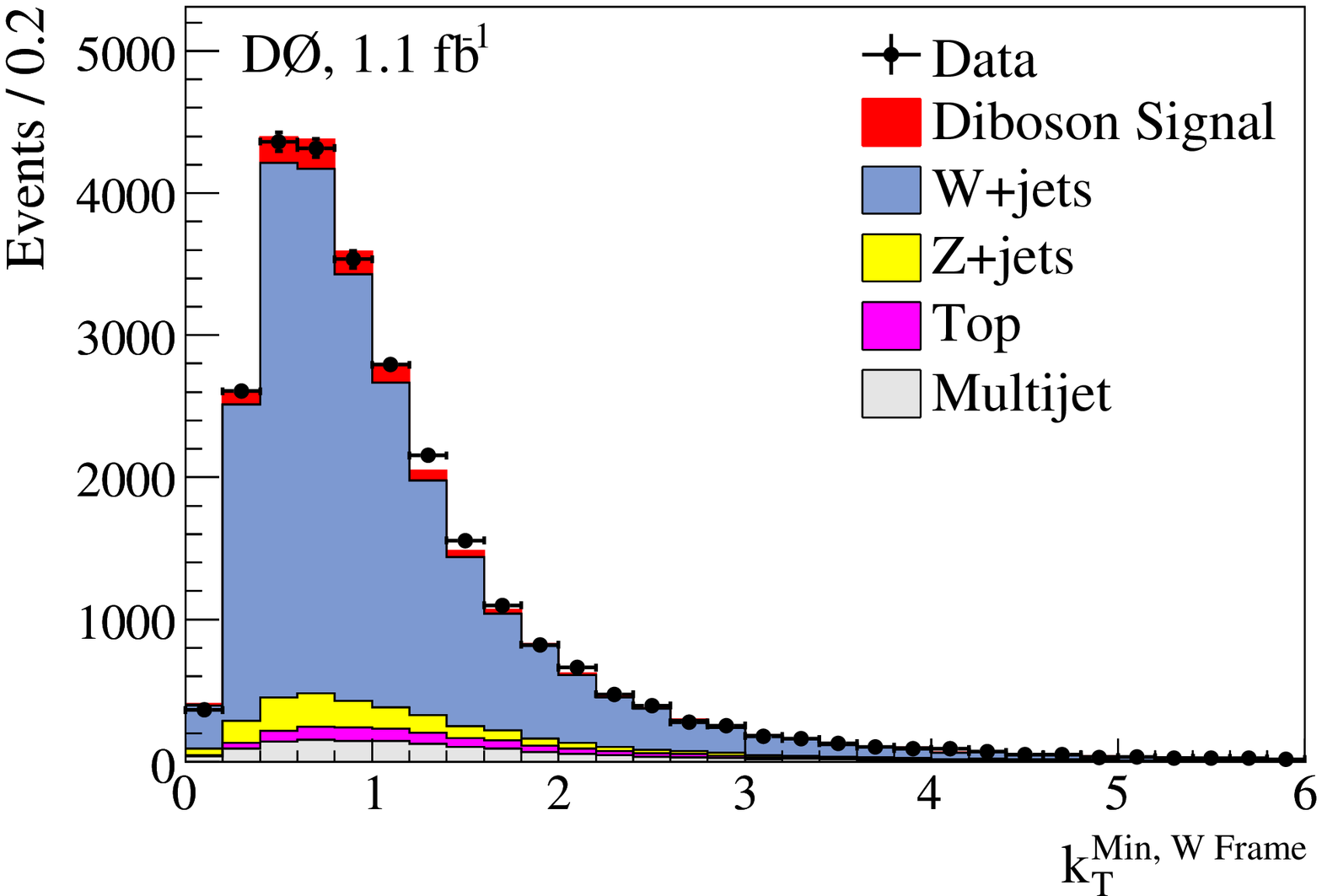}
  \includegraphics[width=5.9cm]{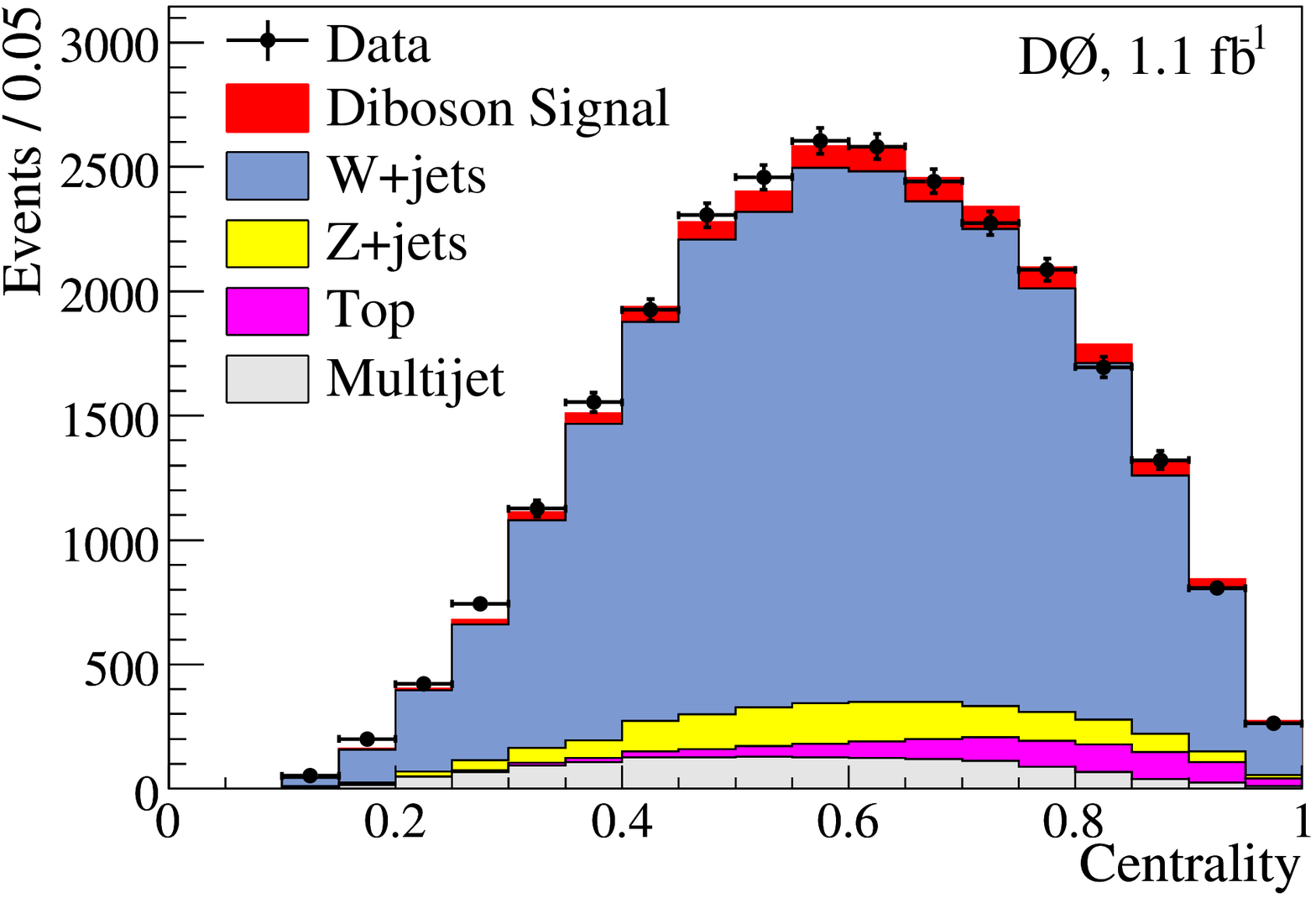}
  \includegraphics[width=5.9cm]{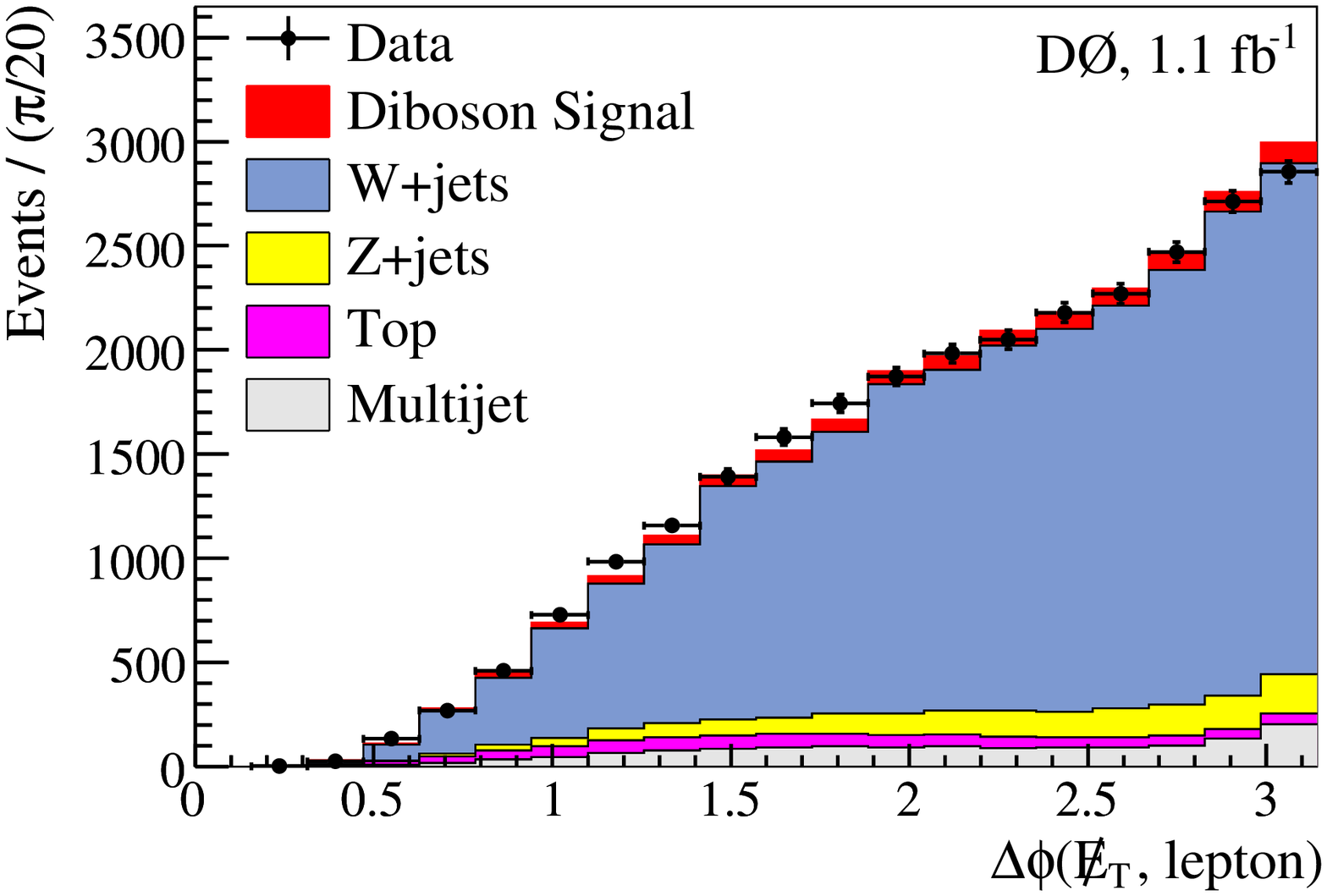}
  \includegraphics[width=5.9cm]{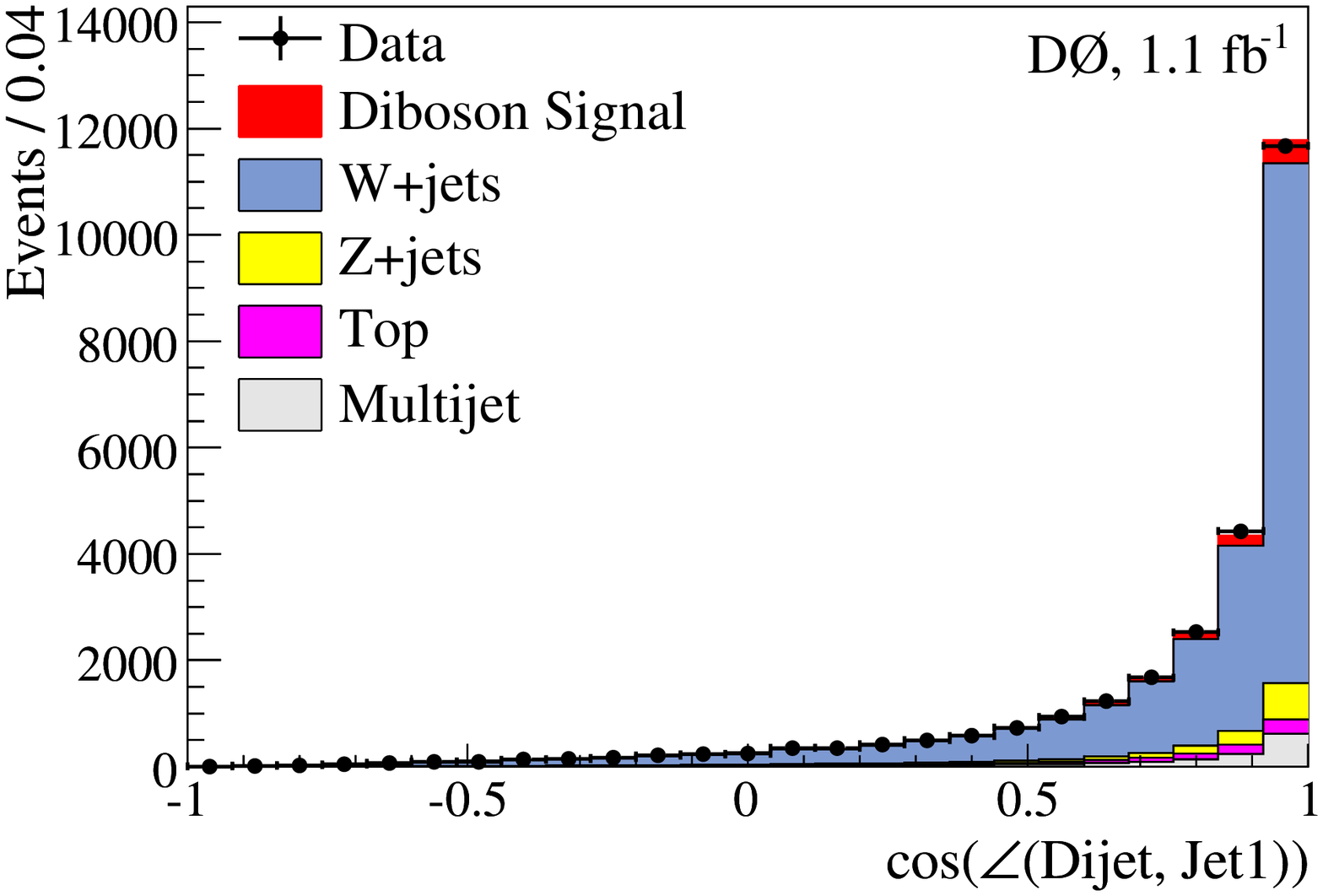}
  \includegraphics[width=5.9cm]{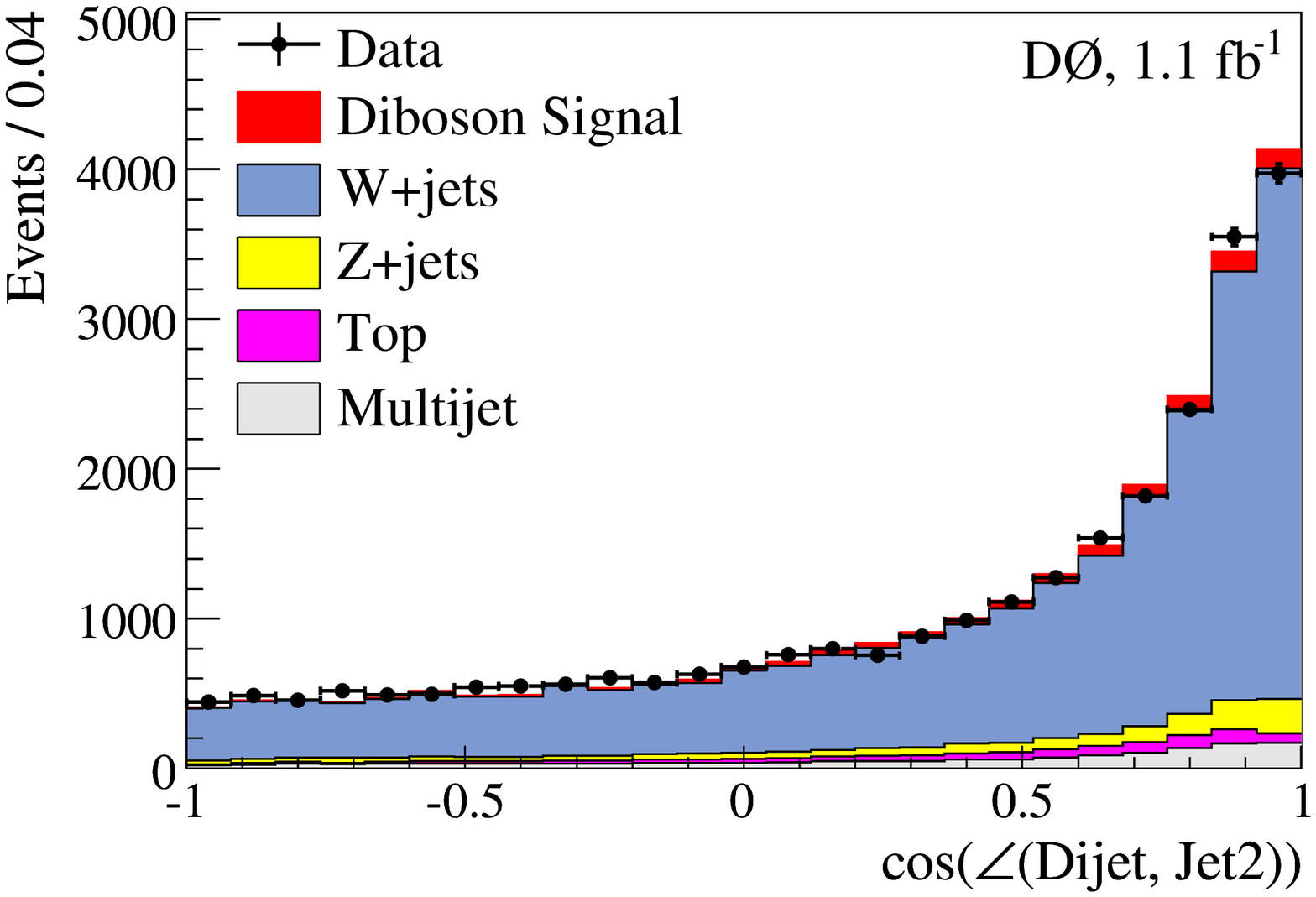}
  \includegraphics[width=5.9cm]{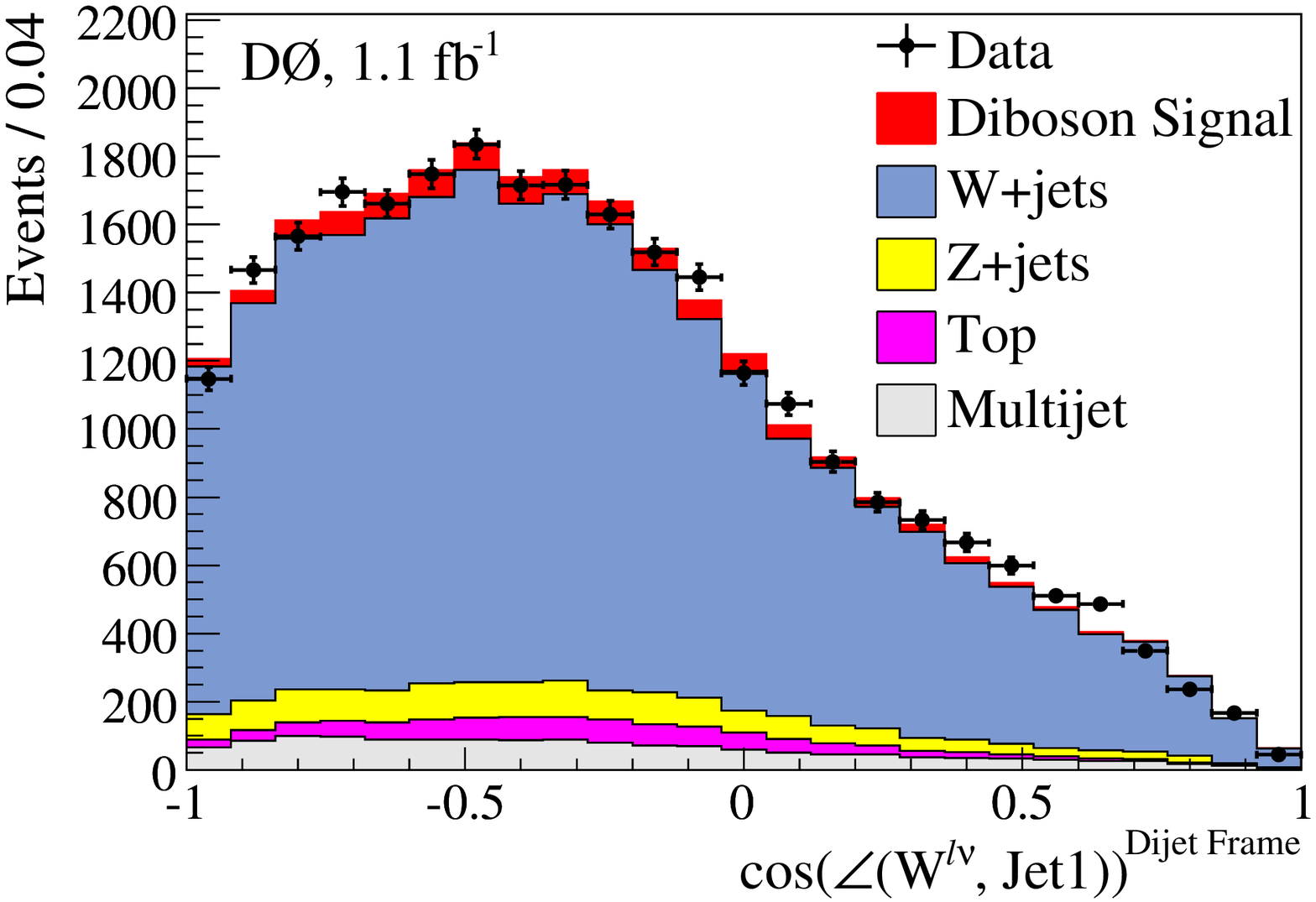}
  \caption{Distributions of the RF input variables for the combined
  $e\nu q\bar{q}$ and $\mu\nu q\bar{q}$ channels comparing the data
    with the MC predictions, following the fit of MC to data.}
  \label{fig:rf_inputs}
\end{figure*}

\section{Correlation between Dijet Invariant Mass and RF Output}

There is a high degree of correlation between the dijet invariant mass
and the RF output.  This can be observed in the dijet invariant mass
distributions for events with low, intermediate and high values for
the RF output shown in Fig.~\ref{fig:mjj_regions}.  As expected,
events in the low region of the RF output correspond to the background
dominated sidebands of the dijet invariant mass distribution and
events in the high region of the RF output correspond to the signal
resonance region of the dijet invariant mass.  The purity of the
signal in the dijet invariant mass distribution is enhanced for high
values of the RF output because a substantial fraction of the
background events in the dijet invariant mass signal region has been
moved to the intermediate region of the RF output.

\begin{figure*}[htb]
  \centering
  \begin{tabular}{ccc}
    \includegraphics[width=5.9cm]{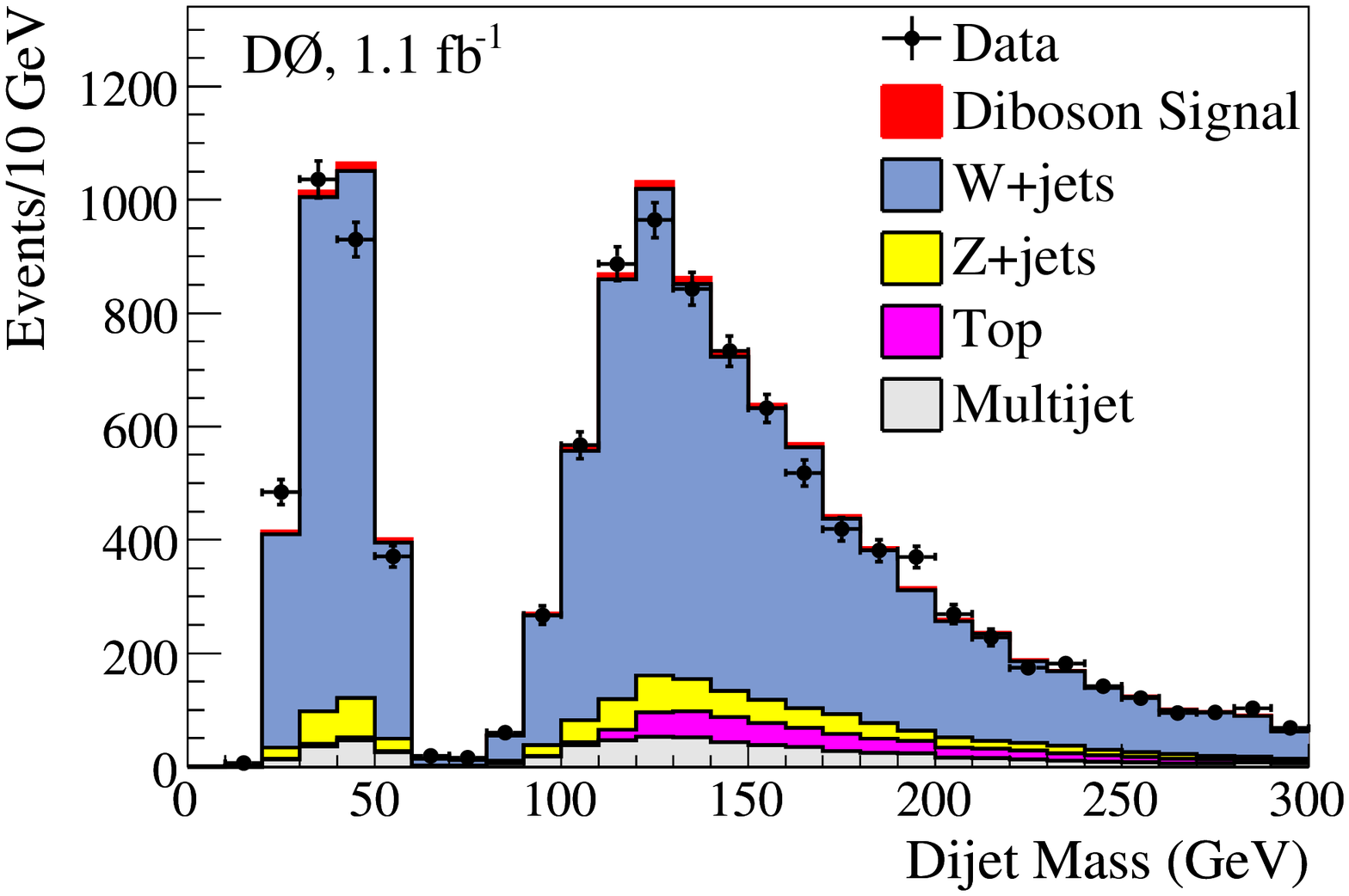}&
    \includegraphics[width=5.9cm]{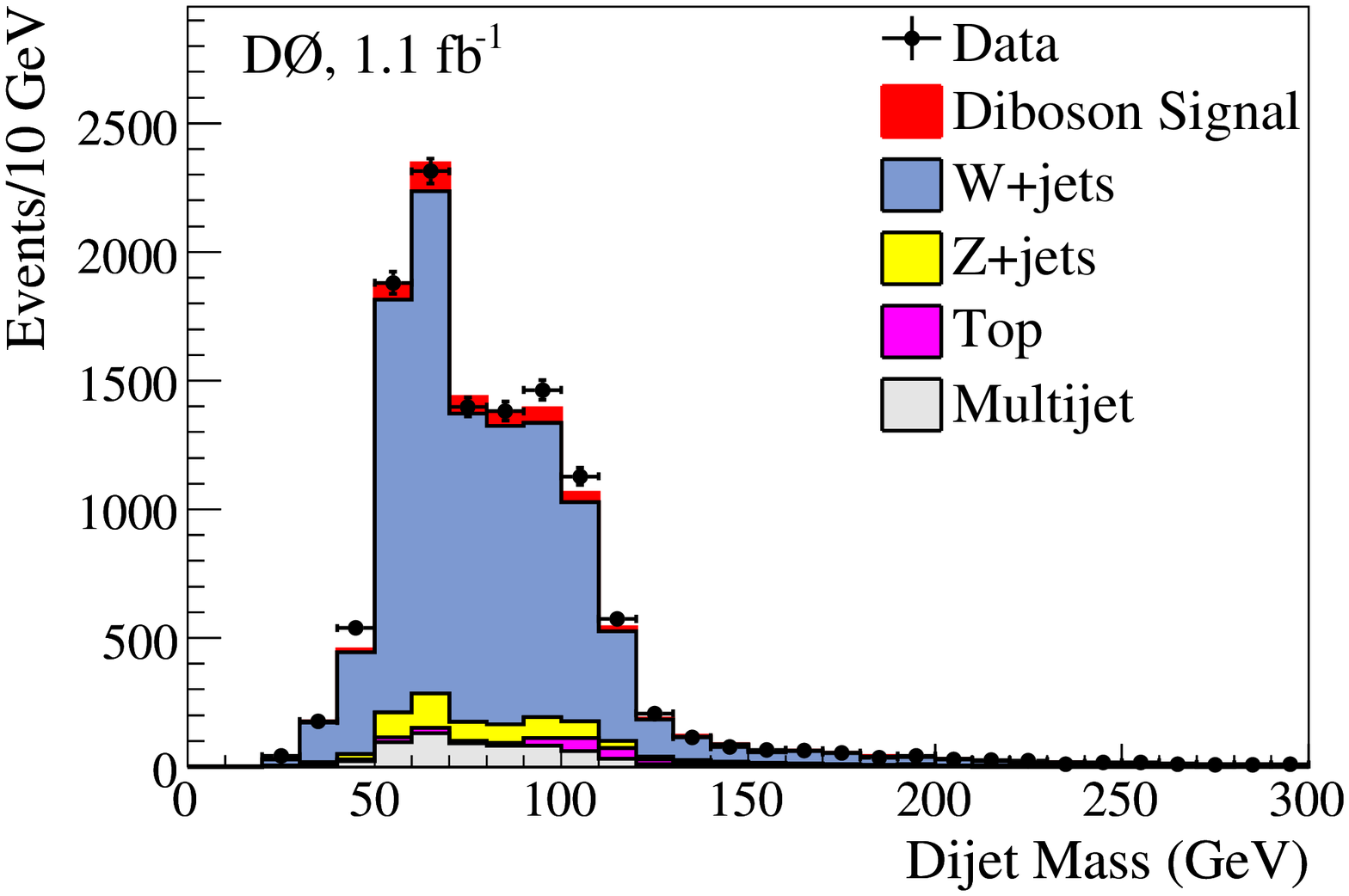}&
    \includegraphics[width=5.9cm]{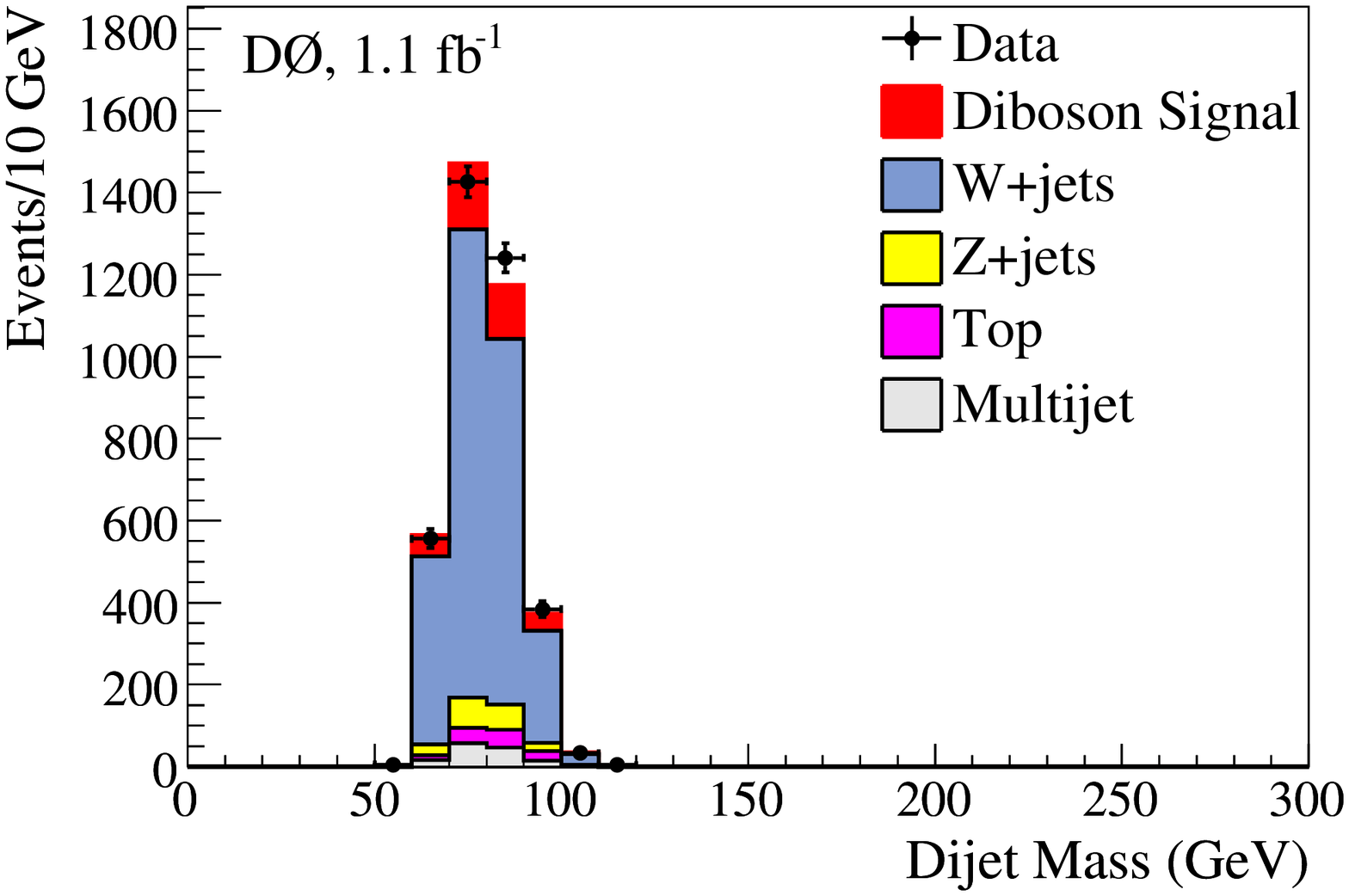}\\
    (a)&
    (b)&
    (c)\\
  \end{tabular}
  \caption{Distributions of the dijet invariant mass for the combined $e\nu
    q\bar{q}$ and $\mu\nu q\bar{q}$ channels comparing the data with
    the MC predictions for events in three regions of the RF output:
    (a) $0 \le$ RF output $\le 0.33$, (b) $0.33 <$ RF output $\le
    0.66$ and (c) $0.66 <$ RF output $\le 1$.}
  \label{fig:mjj_regions}
\end{figure*}

\clearpage


%

\end{document}